# Constants of extended Standard Model and search for their temporal variations [*]

*S.A. Kononogov, V.N. Melnikov, V.V. Khruschov*

Some methods for the determination and the most precise values of constants of an extended Standard Model, which includes the gravitation interaction and massive neutrinos, are presented. Accuracies of constants at different energy scales are compared and the possible manifestations of temporal variations of constants are considered. Theoretical estimations and obtained experimental bounds of variations are listed, that is important as for search of SM generalizations as to account for possible influence on metrological characteristics of measurement standards.

In spite of its success the Standard Model (SM) is a intermediate model for a creation of a more elaborate theory. We have many SM constants together with an absence of an explanation of values and possible links between them. Besides that experimental data exist in favour of a SM generalization such as neutrino oscillations and nonzero neutrino masses, discoveries of dark mass and dark energy [1, 2]. These facts lead to a extension of the SM with a gravitation theory and an inclusion of neutrino masses.

Let us list constants of the extended SM (ESM) among them constants of a neutrino sector and the gravitation constant $G$ [1-5]. The coupling constants of the strong, electromagnetic and weak interactions are $\alpha_c$, $\alpha$ and $\alpha_w$. The QCD vacuum phase, which breaks down the CP invariance, is negligible ($< 10^{-9}$), so this value is set equal to zero. Mass parameters of the ESM are the six quark masses, the six lepton masses, and the Z boson mass. Mixing parameters are the four quark mixing parameters and the six neutrino mixing parameters. At present there are restrictions for the Higgs boson mass at 95% CL: *114,4 ГэВ ≤ $M_H$ ≤ 160* [6, 7]. Thus the ESM contains the 29 parameters or constants. Of this list of constants the Higgs boson mass, a neutrino mass and two Majorana mixing phases are not measured.

In recent years possible time and space variations of ESM parameters (or fundamental physical constants (FPC)) are being studied intensively as in the framework of Great Unification Theories (GUTs) as with phenomenological models [8-25]. Observations produce several restrictions on time variations of FPC, they are connected with Bing Bang nucleosynthesis, quasar spectra, laboratory experiments and Oklo natural nuclear reactor data.

Verification of time α variations is of primary importance, but it should be taken into account that α variations are connected with variations of other coupling constants and masses of fundamental particles.

### 1. Accuracies of ESM constants at different energy scales

The fine structure constant is the most employed coupling constant, which in the Gauss units has the form $\alpha = e^2/(\hbar c)$, while in the SI units $\alpha = e^2/(4\pi\varepsilon_0 \hbar c)$. QED observables can be evaluated and measured with the highest accuracies. For instance, the energy difference between $1S_{1/2}$ and $2S_{1/2}$ levels of the hydrogen atom is known with the relative uncertainty $1.9\times10^{-14}$ [26], relative uncertainties of frequencies for transitions between different states of caesium and mercury atoms can be determined at the level $10^{-16}$ [27].

In QCD framework a dependence of $\alpha_c = g_c^2/4\pi$ on a transferred momentum ($g_c$ is a coupling constant between color quarks and gluons) can be evaluated with the perturbation theory and a renormalization scheme thereafter can be tested experimentally for transferred momenta larger than a QCD scale constant $\Lambda_{QCD} \equiv \Lambda_c$. The $\Lambda_c$ value is about several hundreds MeVs and depends on a number of active quarks participating in a process [1, 28, 29]. Systematical errors, which have mainly a theoretical origin, contribute significantly to a determination of $\alpha_c$. A function $\alpha_c(\mu)$ of renormalization scale $\mu$ is specified by an equation:

$$\mu d\alpha_c(\mu)/d\mu = 2\beta(\alpha_c) = -\beta_0/2\pi \cdot \alpha_c^2 - \beta_1/4\pi^2 \cdot \alpha_c^3 - \beta_2/64\pi^3 \cdot \alpha_c^4 - \ldots, \quad (1)$$

where coefficients $\beta_i$, $i = 0, 1, \ldots$, of the β- function are the following: $\beta_0 = 11 - 2n_f/3$, $\beta_1 = 51-19n_f/3$, $\beta_2 = 2857 - 5033n_f/9 + 325n_f^2/27$, $n_f$ is a number of active quarks participating in a process at $m_{n_f} < \mu$. From this it follows, that in a one-loop aproximation the function $\alpha_c$ of transferred momentum squared $Q^2$ is:

$$\alpha_c(Q^2) = \alpha_c(\mu^2)/(1+\alpha_c(\mu^2)\beta_0 ln(Q^2/\mu^2)/4\pi) \quad (2)$$

The Eq.(2) for the "running" coupling constant take place in a gauge theory and in doing so in the QED: $\beta_0 = -4/3$, in the GWS theory: $\beta_0 = 10/3$, at $n_f = 6$, in the QCD: $\beta_0 = 7$, at $n_f = 6$. Another parameter $\Lambda_c$ is used in the QCD instead of μ: $\Lambda_c^2 = \mu^2 exp(-4\pi/\beta_0\alpha_c(\mu^2))$. Hence the QED coupling constant α grows when $Q^2$ increases, while $\alpha_w$ and $\alpha_c$ diminish and become small in an asymptotical freedom regime. The Fermi constant $G_F$ associated with $\alpha_w$ is determined with the μ-meson life-time data $G_F = 1.16637(1)\times10^{-5}$ GeV$^{-2}$. The Z-boson mass value is determined from a shape of the peak line observed at LEP1, $M_Z = 91.1876(21)$ ГэВ [1, 26].

The commonly accepted scale for SM coupling constants is the Z-boson mass. The $\alpha_c$ value is determined with the best precision (~$1.1\cdot10^2$) from τ-meson decay data [1, 29]:

$$\alpha_c (M_Z) = 0.1212 \pm 0.0011 \qquad (3)$$

The value of the electromagnetic coupling at low energies (the fine structure constant α or the Sommerfeld constant), at present is most accurately determined from the measurement of the electron anomalous magnetic moment $a_e$ [30] (with an accuracy ~$1.4 \cdot 10^9$) [31]:

$$\alpha^{-1} = 137.035999710(96) \qquad (4)$$

Using data for the electron-positron annihilation in the processes $e^+ e^- \to \mu^+ \mu^-$ and $e^+ e^- \to e^+ e^- \mu^+ \mu^-$, observed at the TRISTAN accelerator of KEK (Japan), it is obtained with an accuracy ~$10^2$ at 57.77 GeV scale [32]:

$$\alpha^{-1} = 128.5(2.5) \qquad (5)$$

The α value increases still further when the energy grows and at the Z-boson mass scale is (with an accuracy ~$7.1 \cdot 10^3$) [3]

$$\alpha^{-1} = 127.918(18) \qquad (6)$$

Note that these α variations depending on an energy scale manifest themselves at energies which are higher then energies of atomic transitions, i.e. energy scale α variations do not directly relate with very slow time variations considered below (see, e.g., [17]).

The accuracy of determination of SM physical observables is very high in measurements of lepton anomalous magnetic moments $a_l$. As it noted above the α value has the highest precision from $a_e$ measurement data. Notice that there is the discrepancy between the predicted and experimental value of muon magnetic moments at the 3,4 σ level, that indicates possible a contribution of "new physics" effects [33]. Moreover there is the intriguing difference (~ 3,2 σ) between the values of the effective Weinberg angle measured from lepton and hadron contributions: $\sin^2(\theta_{eff})_l = 0.23113(21)$, $\sin^2(\theta_{eff})_h = 0.23222(27)$ [33].

It is known, that quark masses cannot be measured in free states or any macroscopic external field, they have not classical limits unlike, for example, the electron mass [34]. In the QCD one cannot put a quark mass value at some "natural", "physical" scale. Quark masses are not uniquely defined, they depend on a renormalization scheme and their values $m(\mu)$ are governed by the following equation [1, 5]:

$$\mu^2 dm(\mu)/d\mu^2 = -\gamma(\alpha_c(\mu))m(\mu), \qquad (7)$$

where γ is an anomalous dimension, that is known in the perturbation theory up to fourth order, for instance,

$$\gamma(\alpha_c(\mu)) = \sum_{r=1}^{\infty} \gamma_r (\alpha_c/4\pi)^r, \quad \gamma_1 = 4, \quad \gamma_2 = 202/3 - 20 n_f/9, \quad \ldots, \qquad (8)$$

$n_f$ is a number of active quarks participating in a process at $m_{n_f} < \mu$. So quarks become lighter with increasing the transferred momentum. The quark mass values in the $\overline{MS}$-renormalization scheme at the 2 GeV scale for light quarks and at the quark mass scale for heavy quarks are the following: $m_u$ = 1.5÷3 MeV, $m_d$ = 3÷7 MeV, $m_s$ = 70÷120 MeV, $m_c$ = 1.16÷1.34 GeV, $m_b$ = 4.13÷4.27 GeV [1].

The well elaborated procedure exists for determination of the strong coupling constant at large transferred momenta but contributions of threshold effects should be taken into account because of a quark production. The $\Lambda_c$ value depends on a number of active quarks $n_f$, i.e. $\Lambda_c \rightarrow \Lambda_c(n_f)$, the $\Lambda_c(n_f)$ value decreases if $n_f$ increase, at $n_f$ = 5 $\Lambda_c^{(5)}$ = 217±25 MeV [1]. When momentum values are of order of magnitude $\Lambda_c$, it is necessary to take into account nonperturbative contributions, which are dominating at $Q^2 \ll \Lambda_c$. This momentum range is named the QCD nonperturbative region; it turns into the confinement region. The belief is that the confinement of quarks and gluons can be consistently proved in the QCD framework. If it is the case, then dependence exists between the typical phenomenological parameter of strong interactions namely the string tension $\sigma_s$ and the $\Lambda_c$. If the confinement in the QCD framework is not proved, a number of strong interaction parameters will be increased because of an introduction of a new parameter e.g. the $\sigma_s$.

In this connection works concerning with modifications of a space-time symmetry of the QCD in the confinement region are of interest [35]. It is known that constituent quarks exist due to nonperturbative interactions. Energies of constituent quarks and their systematical uncertainties can be estimated in the framework of the relativistic quasi-independent quark model: $E_u$ = 335±2 MeV, $E_d$ = 339±2 MeV, $E_s$ = 485±8 MeV, $E_c$ = 1610±15 MeV, $E_b$ = 4952±20 MeV. The relation between a current quark mass and a constituent quark mass is the following: $M_q(\mu_0) \approx m_q(\mu_0) + \delta_s$, $\delta_s \approx 314$ MeV, $m_q(\mu_0)$ is the current quark mass at the scale $\mu_0$ [35], where $\mu_0$ is a characteristic quark momentum in a ground hadron state, i.e. $\mu_0 \approx$ 1GeV. Thus

$m_u(\mu_0)$ = 2,2±2 MeV, $m_d(\mu_0)$ = 6,3±2 MeV, $m_s(\mu_0)$ = 158±8 MeV,

$m_c(\mu_0)$ = 1292±15 MeV, $m_b(\mu_0)$ = 4637±20 MeV. (9)

These values are in agreement within 3σ limits with the current quark masses displayed above which have been obtained from deep inelastic processes. The t quark mass cannot be determined analogously, because of a non-existence of bound states with t quarks, but $m_t$ = 172.6±1.4 GeV from t quark production processes [36].

At present the Cabibbo-Kobayashi-Maskava matrix elements are precisely known. In the standard parameterization for the CKM matrix we have [1, 37]

$\delta = 63°(+15°/-12°)$, $\theta_{12} = 13.14°\pm0.06°$, $\theta_{23} = 2.43°(+0.01°/-0.05°)$, $\theta_{13} = 0.23°\pm0.01°$ (10)

Angles and phases for the Pontecorvo-Maki-Nakagava-Sakata matrix are not all known in the lepton sector. The angle $\theta_{sol} = 34.01°(+1.31°/-1.56°)$ is known with the ~2·10 accuracy, while the restrictions exist for $\theta_{atm}$ and $\theta_{rea}$ ($\equiv\theta_{chz}$) angles: $\theta_{atm} > 36.78°$, $\theta_{rea} < 12.92°$. The most difficult task is a determination of Dirac and Majorana mixing phases for neutrinos. Now it is obtained that the two neutrino masses have nonzero values: $\Delta m^2_{sol} = (8.0\pm0.3)\cdot10^{-5}$ $eV^2$, $|\Delta m^2_{atm}| = (2.45\pm0.55)\cdot10^{-3}$ $eV^2$. Schemes exist, in frameworks of which relations and values of quark and neutrino mixing angles can be obtained. For example, one can evaluate quark and neutrino mixing angles with the help of constituent quark masses. These angles are in accordance with experimental data and $\theta_{atm}$ and $\theta_{rea}$ have more precise uncertainties as compared to data [37]. At 90% CL these values are the following:

$\theta_{sol} = 31.9°\pm2.8°$, $\theta_{atm} = 42.53°\pm0.05°$, $\theta_{rea} = 0.206°\pm0.005°$ (11)

Note that the precision of the gravitation constant value is not very high. In 1998 the level of relative standard uncertainties was reduced from $10^{-4}$ to $10^{-3}$ on account of discrepancies among several experimental groups. At present the precision of the G is $1\cdot10^4$ [1, 26]

$G = 6,67428(67)\cdot10^{-11}$ $m^3 kg^{-1} s^{-2}$ (12)

## 2. Theoretical estimations of possible time variations of ESM constants

There are several problems in the SM, which can be solved in a more general theory. For instance, in the GUT framework the $SU(3)_c \times SU(2)_L \times U(1)_Y$ gauge group changes to more simple group [1, 4, 23]. The pioneering GUTs were based on $SU(4)_{ec} \times SU(2)_L \times SU(2)_R$ [38], $SU(5)$ [39] and $SO(10)$ [40, 41] groups. Today the $SO(10)$ model is of considerable interest with the following scheme of spontaneous breaking of symmetry:

$SO(10) \rightarrow SU(4)_{ec} \times SU(2)_L \times SU(2)_R \rightarrow SU(3)_c \times SU(2)_L \times U(1)_Y$ (13)

In the SO(10) model the minimal spinor 16-dimensional representation consists of left fundamental fermions, i.e. quarks and leptons including CP partner of a right neutrino. So it is easy to incorporate a massive neutrino in the SO(10) model. Moreover this model does not contradict existing experimental data concerning to the Weinberg angle, the proton life time, etc [1]. Supersymmetric GUTs are of interest as well because of a concordance with data and a coincidence of coupling constants at an energy of unification in their frameworks.

In searching for temporal variations of strong, electromagnetic and weak coupling constants a joint evolution of all constants must be taken into consideration [23]. Really in a GUT at superhigh energies of interactions ($\sim 10^{16}$ GeV) the coupling constants close together and give to a coupling constant of an universal interaction [1, 3-5], so variations of all coupling constants at large scale are related with each other and should be jointly be taken into account. Let us consider estimations of temporal variations of coupling constants in the framework of the minimal supersymmetric standard model (MSSM) [42]. If a energy of unification is $E_G$, then

$$\alpha(E_G) = \alpha_w(E_G) = \alpha_s(E_G) = \alpha_G \tag{14}$$

In accordance with renormalization group equations constants $\alpha_i = g_i^2/4\pi$, $i = 1, 2, 3$, which are connected with the $U(1)xSU(2)xSU(3)$ group, obey the equation (1) at a variation of a renormalization point $\mu$ like it is in the QCD. Thus the equation (2) for the running coupling constants $\alpha_i$ will be valid in any gauge theory. Let us take $\mu_G = E_G$, then the equation (2) can be written in the form:

$$\alpha_i^{-1}(Q) = \alpha_G^{-1} + \beta_{0i} l_G, \tag{15}$$

where $l_G = ln(\mu_G/Q)/2\pi$. It is known coefficients $\beta_{0i}$ depend on a set of gauge fields and fundamental fermions. In the MSSM framework they are $\{\beta_{0i}, i = 1, 2, 3\} = \{33/5, 1, -3\}$. So the strong coupling constant $\alpha_c \equiv \alpha_3$ decreases and becomes small in the asymptotic freedom region, while the constants $\alpha_1$ and $\alpha_2$ connected with the electrodynamics and weak interactions grow with increasing of $Q^2$ values.

The Eq. (15) is usually used in order to determine a time evolution of coupling constants in the one-loop approximation. In doing so it is assumed that the term $\alpha_G^{-1}$ gives the main contribution, whereas contributions of the terms $\beta_{0i} l_G$, $i = 1, 2, 3$, can be neglected. At a low energy limit variations of constants are compared at the common scale equal to the Z boson mass and variations at scales lower than the Z boson mass are not considered. Take into account that in the MSSM model $\alpha^{-1}(M_Z) = (5/3)\alpha_1^{-1}(M_Z) + \alpha_2^{-1}(M_Z)$, the following estimations can be obtained [42].

$$\Delta\alpha_i/\alpha_i = -\Delta\alpha_i^{-1}/\alpha_i^{-1} = -\Delta\alpha_G^{-1}/\alpha_i^{-1},$$
$$\Delta\alpha/\alpha = -8\Delta\alpha_G^{-1}/3\alpha_i \approx 0.49 \Delta\alpha_G/\alpha_G, \tag{16}$$
$$\Delta\alpha_c/\alpha_c \approx 5.8 \Delta\alpha/\alpha, \quad \Delta\Lambda_c/\Lambda_c = (2\pi/9\alpha_c) \Delta\alpha_c/\alpha_c, \quad \Delta\Lambda_c/\Lambda_c \approx 34 \Delta\alpha/\alpha$$

The accuracy of these estimations is about 20%. The last two estimates of the set (16) is important because terms proportional to $\Lambda_c$ provide the main contribution in mass values of protons, neutrons and nuclei. However masses of leptons and quarks are determined with a Yukawa interaction of a Higgs field with fundamental fermions: $m_a = y_a v$, where $y_a$ is the Yukawa coupling constant, $v$ is the scale of electroweak interactions (approximately equal to the 247 GeV). At present the Higgs boson is not discovered, so assumptions underlying the structure of the Higgs sector of the SM are not justified. Hence the problem of time variations of quark and lepton masses is more complicated than the problem of time variations of coupling constants. Nonetheless one can obtain the written below estimations of relative variations of quark masses using the usual SM structure of the Higgs sector [42, 18]

$$\delta m_q/m_q \approx 70\, \delta\alpha/\alpha, \quad \delta(m_q/\Lambda_c)/(m_q/\Lambda_c) \approx 35\, \delta\alpha/\alpha \quad (17)$$

These estimations demonstrate the enhancement of time variations of constants in strong interaction effects. The account of strong interaction effects is important for interpretations of data concerning to the anomalous magnetic muon moment [43, 33], the hyperfine splitting of positronium energy levels [44] and the Oklo natural nuclear reactor [45].

As is known assumptions of possible time variations of constants were first connected in the early last century with variations of the light velocity or the gravitational constant. The Dirac hypothesis of large numbers is the most famous one. Under this hypothesis, any physical large number ($\sim 10^{20}$, $\sim 10^{40}$, $\sim 10^{80}$) is correlated with the dimensionless age of the Universe T, which is the ratio of the Universe age $\sim 10^{17}$s and the characteristic life-time of a strong interaction particle $\sim 10^{-23}$ s. Dirac supposed that atomic constants were stable while the gravitational constant could vary according to the law: $G \sim T^{-1}$. Now it is known relative time variations of the gravitational constant are bounded at least with the constraint $|G'/G| < 10^{-12}$ year$^{-1}$ [12, 20].

3. **Restrictions on possible time variations of ESM constants from astrophysical data**

Observations of spectra of astrophysical sources present a possibility to obtain data for energy levels at instants when radiation leaves atoms. However many agents exist, which cause systematical uncertainties and reduce the accuracy of measurement of atomic constants [12]. It is known some observations indicate increasing of $\alpha$ with time. Quasar spectra observed with the Hawaii telescope at 0.2 < z < 3.7 suggest that [46]

$$\Delta\alpha/\alpha = (-0.57 \pm 0.11)\cdot 10^{-5}. \quad (18)$$

The data result in the following relative variation of $\alpha$ during year, when $\dot\alpha/\alpha$ is unalterable for all z values [46, 8, 47]:

$$\dot\alpha/\alpha \approx 1\cdot 10^{-15}\, \mathrm{yr}^{-1} \quad (19)$$

However some observations are not in line with variations of $\alpha$ in astrophysical spectra. For instance, it has been obtained in absorption quasar spectra at z = 1.15 [48]

$$\Delta\alpha/\alpha = (-0.1 \pm 1.7) \cdot 10^{-6} \qquad (20)$$

Similar data have been obtained in absorption spectra of Q 1101-264 at z = 1.839 [49]

$$\Delta\alpha/\alpha = (2.4 \pm 3.8) \cdot 10^{-6} \qquad (21)$$

Moreover study of absorption quasar spectra in radio and millimeter range at 0.25 < z < 0.68 [50] leads to

$$|\Delta\alpha/\alpha| < 8.5 \cdot 10^{-6} \qquad (22)$$

A thorough analysis has been done [51], which confirmed that the results of $\alpha$ variations in spectra of quasars are contradictory and a revision of results obtained in the southern hemisphere is possibly required.

Now indications were also received of possible changes over time for ratio of proton and electron masses $\mu = m_p/m_e$ in addition to results associated with $\alpha$ variations. For example, it has been find the variation of $\mu$ during $10^{10}$ years in the Q 0347-382 quasar spectrum with the help of the method based on dependence on $\mu$ for lengths of waves generated due to transitions between oscillatory-rotational levels of molecules [11].

$$\Delta\mu/\mu = (5.02 \pm 1.82) \cdot 10^{-5} \qquad (23)$$

Observations of spectral lines of $H_2$ in the Q 0347-383 and Q 0405-443 quasars lead to the $\mu$ variation during $1.2 \cdot 10^{10}$ years as well [19, 21]:

$$\Delta\mu/\mu = (2.4 \pm 0.6) \cdot 10^{-5} \qquad (24)$$

It gives the following value:

$$\dot{\mu}/\mu \approx -2.0 \cdot 10^{-15} \text{ yr}^{-1} \qquad (25)$$

Values of relative $\alpha$ and $\mu$ variations have different signs that contradicts estimations of their variations in the SM framework. But some models are available, for which such variations are permitted [21].

The analysis of astrophysical data for superdense stars [52] leads to the following restrictions on the temporal variations per year for the $\Lambda_c$ and the gravitational constant:

$$|\dot{\Lambda}_c/\Lambda_c| < 10^{-17} \div 10^{-15} \text{ yr}^{-1} \quad ,$$
$$G'/G < 10^{-17} \div 10^{-15} \text{ yr}^{-1} \qquad (26)$$

More conservative estimates derived from data with laser-scanning of the Moon [53] lead to the result

$$G'/G = (4 \pm 9) \cdot 10^{-13} \text{ yr}^{-1} \qquad (27)$$

Despite the fact that the situation in general with the astrophysical data on temporary variations FPC remains controversial, it is desirable to continue to pursue both theoretical and experimental research in this direction. It may be necessary to refuse the relative constancy of the speed of FPC variations and to use an assumption of the nonlinear FPC evolution, provided that the most rapid changes of constants take place in the early Universe. In doing so, the restrictions on changes of $\alpha$ and other constants must take into account that emerges from data on production of elements during the primary nucleosynthesis [23].

4. **Restrictions on possible time variations of ESM constants from laboratory data**

For monotone temporary FPC variations one should compare their values with existing uncertainties of constants. Let us present uncertainties for most interesting constants such as $\alpha$ and the electron mass $m_e$. The $\alpha$ value is most accurately determined from the data on measurements of the anomalous magnetic moment of the electron. In the Dirac theory g-factor equals to two, the QED predicts deviation from this value due to emission and absorption of virtual photons, thus increasing g and there arise a nonzero anomalous magnetic moment of the electron: $a_e = (g - 2)/2$. Now an accuracy of the experimental $a_e$ determination is very high, the standard uncertainty is $0.76 \cdot 10^{-12}$. A theoretical forecast for $a_e$ value is weakly sensitive to the weak and strong interactions and is the most rigorous QED test. Sharing experimental and theoretical results [30, 31, 54] leads to the most precise $\alpha$ determination with the standard relative uncertainty $\sim 0.7 \cdot 10^{-9}$. An alternative method for the $\alpha$ determination on the same level of accuracy can be a comparison of calculated and experimental values of g-factors of various lead ions [55].

The electron mass is defined with the greatest accuracy in ion Penning traps through an indirect measurement procedure, in which the results of calculations carried out within the QED framework are played the major role. For it characteristics of motion and radiation of captured ions are experimentally determined, followed by comparison with a theoretical value of g-factor of a bound electron, which includes electromagnetic corrections as well as corrections due to a motion and finite size of nucleus [56, 57]. Experiments have been performed for ions of oxygen and carbon captured in the Penning trap. The Penning trap is suitable for $m_e$ determination with the best accuracy in the atomic mass units through comparison of cyclotron frequencies of electrons and ions captured in a trap [34, 58, 59]. In the near future ratios of Larmor and cyclotron frequencies for hydrogen like ions can be measured in experiments with Penning traps with relative uncertainties $\sim 10^{-11}$ [60].

Let us show values of lepton masses obtained to date: $m_e$ = 0.510998910(13) MeV {4·10$^6$}, $m_\mu$ = 105.6583692(94) MeV {1.1·10$^7$}, $m_\tau$ = 1776.90(20) MeV {9·10$^3$} [1, 26] (the precision of measurement is displayed in the braces). As it is known the accuracy of the determination of physical quantity depends on several factors, including the selected unit of measurement. For instance, the values of electron and muon masses in atomic mass units (u) are the following: $m_e$ = 5.4857990945(24)×10$^{-4}$ u {2.3·10$^9$}, $m_\mu$ = 0.1134289264(30) u {3.8·10$^6$}.

Taking into consideration a vague state with the experimental confirmation of time variations of α and μ from astrophysical data, it is important to take into account precise laboratory data. From the hyperfine atom transitions the restrictions have been obtained on time variations of α and $m_p/\Lambda_c$, which are from 10$^{-16}$ to 10$^{-14}$ per year for relative variations [18]. Comparisons of drifts for mercury, rubidium and ytterbium atomic clocks with the cesium standard give an upper limit about 10$^{-15}$ per year [61, 18]. Recently, for the first time the comparison of frequencies for the SF$_6$ molecular and cesium clocks has been performed [25], which allows to check with high precision the μ stability.

$$\dot{\mu}/\mu \approx (-3.8 \pm 5.6) \cdot 10^{-14} \, \text{yr}^{-1}. \qquad (28)$$

A recently calculated estimation for the lower limit of $m_q/\Lambda_c$ time variation, which can be achieved in the $^{229}$Th experiment [62], show that this limit can be by several orders of magnitude greater than existing restrictions of such variations, obtained with atomic clocks.

**5. Restrictions on possible time variations of ESM constants from Oklo natural nuclear reactor data**

It is known, the most stringent restrictions on the α time variations were obtained from the study of the mixture of chemical elements nearby the natural nuclear reactor at the Oklo River. The ratio of $^{149}$Sm and $^{147}$Sm isotopes is measured. This ratio is about one under normal conditions, while in the Oklo area it is two orders of magnitude less. The satisfactory explanation serves as the existence of the transition from $^{149}$Sm to $^{150}$Sm due to the neutron irradiation:

$$^{149}\text{Sm} + n \rightarrow {}^{150}\text{Sm} + \gamma \qquad (29)$$

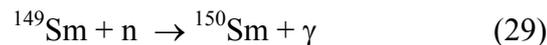

The reaction enhances by presence of $E_r \cong 0.0973$ eV resonance, whose position could not change by more than 0.1 eV over the past ~ 1.8 x10$^9$ years [63, 12].

In all investigations until 2002 only the α variation was taken into account, resulting in the restriction $\Delta\alpha/\alpha < 10^{-7}$ over ~ 1.8 x10$^9$ years and provided that the relative velocity of changes remains constant throughout this period. However, as noted above, in grand unification models the α temporal variations must be accompanied by variations of other constants primarily strong interaction constants [12, 42, 64]. If you consider that the temporal variations of constants of strong interactions are associated only with $\Lambda_c$, then the result obtained above practically does not change, because all nuclear parameters undergo the same shift. But taking into account, that current quark masses are more sensitive to time variations [42, 18]:

$$\delta m_q / m_q \approx 70\ \delta\alpha/\alpha, \quad \delta(m_q/\Lambda_c)/(m_q/\Lambda_c) \approx 35\ \delta\alpha/\alpha, \quad (30)$$

one can get the strongest restriction $\Delta\alpha/\alpha < 5 \cdot 10^{-10}$, which allows for the relative velocity of α changes per year: $\dot{\alpha}/\alpha < 2.5 \cdot 10^{-19}$ yr$^{-1}$, while conservative estimates based on a variation of α only give the following limit: $\dot{\alpha}/\alpha < 5 \times 10^{-17}$ yr$^{-1}$ [64].

### 6. Variations of ESM constants and gravitation theories

It is well known, a quantum theory of gravity is not built so far, and gravitational waves are not detected experimentally. The classical theory of gravitation is the general theory of relativity (GTR), which explains practically all known experimental data. The GTR basis is the Einstein-Hilbert action

$$S_{EH} = (16\pi G)^{-1} \int d^4x \sqrt{g}\ R, \quad (31)$$

where R is the Ricci scalar curvature. To include the theory of gravitation in the SM at the phenomenological level [4], it is necessary to restrict itself only weak gravitational fields and to decompose the metric tensor $g_{\alpha\beta}$ nearly the flat space metric tensor $\eta_{\alpha\beta}$:

$$g_{\alpha\beta} = \eta_{\alpha\beta} + \sqrt{G}\ h_{\alpha\beta}, \quad (32)$$

here $h_{\alpha\beta}$ is a boson field normalized in the usual way. The gravitational constant G is inversely proportional to the Planck mass of second-degree ($M_{Planck} \sim 10^{18}$ GeV), so $h_{\alpha\beta}$ coupling with matter fields will lead to a nonrenormalized theory. However, if one limits a tree approximation in a perturbation theory, which contains no loops with gravitons, then we will get a phenomenological quantum theory, agreed with the quantum theories of strong and electroweak interactions. This theory contains a massless graviton and describes all known effects of the classical Newton and Einstein theories of gravity. The SM with massive neutrinos and the phenomenological quantum theory of gravity will be called the extended SM (ESM). Moreover, a cosmological term can be added to the ESM, which will describe the "dark energy".

$$S_{ESM} = \int d^4x (L_{EH} + L_{SM} - \lambda\sqrt{g}), \quad (33)$$

where $L_{SM}$ is the SM Lagrangian density with massive neutrinos.

In connection with considerations of generalized theories of interactions, including gravitation, note that an introduction of fundamental constants dependent on spatial or temporal variables violates basic principles of any theory, in particular the Einstein principle of equivalence, which did not allow a dependence of results of nongravitational experiments on a location in the space-time. But there are some means to circumvent this limitation. For instance, one can enter an "efficient field", when the fundamental constants are depended on this field [65], or enter the global dependence of vacuum expectations of scalar fields, such as the Higgs field, on sufficiently large areas of space-time ("domain structure"), within which the vacuum expectations remain constant [66]. Then the violation of the principle of equivalence will only take place on the boundaries of spatial-temporal areas.

**7. Conclusions and discussion**

Search for possible temporary variations of constants of strong, electromagnetic and weak interactions accomplished on the basis of existing experimental laboratory, geochemical and astrophysical data suggests a special role of the most accurately determined constant - the fine structure constant - $\alpha$. The "conservative" top limit on time variation values of $\alpha$, obtained on the basis of analysis of the content of chemical elements near the Oklo natural nuclear reactor, is about $10^{-17}$ yr$^{-1}$, as for the gravitational constant, while for the $\mu$ ratio of proton and electron masses this limit is about $10^{-15}$ yr$^{-1}$.

Thus, in present time possible slow temporal FPC variations can not significantly affect metrological characteristics of measurement instruments, but the account of possible variations is required in connection with the rapid development of experimental tools of measurement and the increase of their accuracy. On the other hand the discovery of temporary FPC variations will be of fundamental importance for physics and metrology and will allow establishing the nature of FPC values. The registration of these variations will make for the deletion of systematic errors when using the values of the constants over large intervals of time. For example, for one of the new method to determine the new SI unit of mass with the help of "atomic kilogram" [67, 68], the time-scale $\Lambda_c$ variation could represent the main problem. But, as noted earlier, the analysis of astrophysical data for superdense stars leds to a significant limitation on temporal variation of this parameter: $|\dot\Lambda/\Lambda_c| < 10^{-15}$.

The discovery of temporary FPC variations would go beyond the SM. It should be noted that currently received experimental data are related to the new, beyond the SM, phenomena. For instance, in order to explain the nature of discovered "dark matter" and "dark energy" it is necessary to introduce new particles and interactions. Another example is the discovery of neutrino masses and oscillations. These facts are of great interest for a development of a generalized theory, which can be able to explain the origin of SM constants. However if, according to J. Wheeler, in each evolution cycle of the universe constants arose anew in conjunction with physical laws that govern the development of the universe, then the values of at least some constants can not be rationally explained.

The modern system of standards of physical quantities based mainly on stable natural phenomena and FPC values. After a proposed introduction in 2011 of new definitions of kilogram, mole, ampere and kelvin [69] all SI units of physical quantities will be FPS based. Therefore precision knowledge of values of constants at different instants is needed to verify fundamental physical theories and to enlarge practical applications of these theories.

This work is supported by the RFBR grant # 07-02-13614-ofi-ts.